\documentclass[Afour,sagev,times]{sagej}

\usepackage{moreverb,url}

\usepackage[colorlinks,bookmarksopen,bookmarksnumbered,citecolor=red,urlcolor=red]{hyperref}

\newcommand\BibTeX{{\rmfamily B\kern-.05em \textsc{i\kern-.025em b}\kern-.08em
T\kern-.1667em\lower.7ex\hbox{E}\kern-.125emX}}

\usepackage{flushend} 

\usepackage{times}

\usepackage{color}

\usepackage{booktabs}

\usepackage{graphics}

\usepackage{url}

\usepackage{amsfonts}

\usepackage{adjustbox}
\usepackage{enumitem}

\def\volumeyear{2018}
\def\journalname{Accepted in Information Visualization}
\def\volumenumber{}
\def\issuenumber{N/A}

\begin{document}


\runninghead{Z. Cui, S. Badam, A. Yal\c{c}in, N. Elmqvist}

\title{DataSite: Proactive Visual Data Exploration with Computation of Insight-based Recommendations}

\author{Anonymous Authors}

\author{Zhe Cui\affilnum{1}, Sriram Karthik Badam\affilnum{1}, M.\
 Adil Yal\c{c}in\affilnum{2}, and Niklas Elmqvist\affilnum{1}}

\affiliation{\affilnum{1}University of Maryland, College Park, MD, USA\\
\affilnum{2} Keshif LLC, Alexandria, VA, USA}

\corrauth{Zhe Cui, University of Maryland, College Park, MD 20742, USA.}
\email{zcui@umd.edu}

\begin{abstract}
  Effective data analysis ideally requires the analyst to have high expertise as well as high knowledge of the data.
  Even with such familiarity, manually pursuing all potential hypotheses and exploring all possible views is impractical.
  We present DataSite, a proactive visual analytics system where the burden of selecting and executing appropriate computations is shared by an automatic server-side computation engine.
  Salient features identified by these automatic background processes are surfaced as notifications in a feed timeline.
  DataSite effectively turns data analysis into a conversation between analyst and computer, thereby reducing the cognitive load and domain knowledge requirements.
  We validate the system with a user study comparing it to a recent visualization recommendation system, yielding significant improvement, particularly for complex analyses that existing analytics systems do not support well.
\end{abstract}

\keywords{Exploratory analysis, user interfaces, proactive visualization, visual insights.}


\maketitle

\section{Introduction}
\label{sec:introduction}

Data exploration using visual analytics~\cite{Thomas2005} is often characterized as a dialogue between analyst and computer, with each conversational partner providing unique and complementary capabilities~\cite{badam2016timefork}: the analyst provides creativity, experience, and insight,  whereas the computer provides algorithms, computation, and storage.
In practice, however, most current visual analytics systems put the analyst in the driver's seat to guide the analysis.
This one-sided arrangement falls short when the analyst does not know how to best transform or visualize the data, or is simply overwhelmed due to the sheer scale of the dataset or the limited time available for analysis.
A balanced dialogue would share control between the two conversational partners---analyst and computer---in a way that leverages their respective strengths.
Such a \textit{proactive approach} to data analysis would automatically select and execute appropriate computations to inform the analyst's sensemaking process.

In this paper, we present \textsc{DataSite}, a proactive visual analytics system where the user analyzes and visualizes the data while a computation engine simultaneously selects and executes appropriate automatic analyses on the data in the background.
By continuously running all conceivable computations on all combinations of data dimensions, ranked in order of perceived utility for the specific data, DataSite uses brute force to relieve the burden from the analyst of having to know all these analyses.
Any potentially interesting trends and insights unearthed by the computation engine are propagated as status notifications on a \textit{feed view}, similar to posts on a social media feed such as Twitter or Facebook.
We designed this feed view to support different stages of exploration.
Status updates are continuously and dynamically added to the feed as they become available during the exploration.
To provide a quick overview, they are presented with a brief description that can be sorted, filtered, and queried.
To get more details on an individual response without committing to the active path of exploration, we allow the analyst to expand an update to see details in natural language as well as an interactive thumbnail of a representative visualization.
Finally, the user can select an update to bring it to the manual specification panel, allowing for manual exploration.

Our web-based implementation of DataSite consists of a web client interface for multidimensional data exploration as well as a server-side computational engine with a plugin system, allowing new components to be integrated.
The client interface is a shelf-based visualization design environment similar to Tableau (based on Polestar~\cite{polestar}).
The server-side computational engine currently includes common analysis components such as clustering, regression, correlation, dimension reduction, and inferential statistics, but can be further expanded depending on the type of data being loaded into DataSite.
Each computational plugin implements a standardized interface for enumerating and ranking supported algorithms, running an analysis, and returning one or several status updates to the feed view.
Computational tasks are run in a multithreaded, non-blocking fashion on the server, and use rudimentary scheduling based on their perceived utility for the dataset.

While our proactive analytics approach and DataSite prototype are novel, they are part of a greater trend on the use of recommendation engines for visualization (e.g.,~\cite{Saket2017, Wongsuphasawat2017, Wongsuphasawat2016}). 
However, additional empirical evaluation is still needed to understand how (a) mixed-initiative and proactive analytics compares to traditional exploratory analysis, as well as (b) specific approaches to this idea compare to each other.
Towards this end,
we present results from two user studies involving exploratory analysis of unknown data, one that compared DataSite to a Tableau-like visualization system (PoleStar~\cite{polestar}), and one that compared it to a partial-specification visualization recommendation system (Voyager 2~\cite{Wongsuphasawat2017}).
Using DataSite's feed, our participants derived richer, more complex, and subjectively insightful findings compared to when using PoleStar, or even Voyager 2's recommendation feed.
This supports our hypothesis that a true proactive analytics platform such as DataSite can improve coverage and increase complexity of insights compared to reactive or partial-specification approaches.
Beyond the DataSite system, our approach can be applied to other exploratory analysis tools to promote richer exploratory analysis, even for non-experts, analysts pressed for time, or analysts unfamiliar with a dataset before exploration.

\section{Background}

DataSite extends the literature on exploratory visual analysis and visualization recommendation to better aid analysts with data exploration in a proactive manner. 
Here we discuss the state-of-the-art research and inspirations for DataSite.

\subsection{Exploratory Visual Analysis}

Exploratory data analysis (EDA)~\cite{keim2006challenges, tukey1977exploratory} is the canonical user scenario for visualization.
The key characteristic for EDA is that the analyst is not initially familiar with the dataset, and may also be unclear about the goals of the exploration.
The exploratory process involves browsing the data to get an overall understanding, deriving questions from the data, and finally looking for answers.

Efficient data exploration often relies on visual interfaces~\cite{tukey1977exploratory}.
\textit{Dynamic queries}~\cite{shneiderman1994dynamic} is an interaction technique for such interfaces, where users formulate visual queries as a combination of filters.
Writ large, \textit{faceted browsing} allows for creating queries on specific dimensions of the data~\cite{yee2003faceted}.

\subsection{Visual Specification}

Specifying visual representations is one of the key challenges in visualization.
Research efforts here span the spectrum from programming languages to point-and-click interfaces.
Visualization toolkits such as D3~\cite{bostock2011d3} represent one side of this spectrum, and gives unprecedented control over the visualization, but at the cost of significant programming expertise and development time.
High-level visual grammars, such as Grammar of Graphics~\cite{wilkinson2006grammar}, ggplot2~\cite{wickham2016ggplot2}, and Vega-Lite~\cite{Satyanarayan2016_vega}, abstract away implementation details, but may still have a high barrier of entry and steep learning curve due to the need for visual design knowledge.

A recent development in visual specification has been the introduction of interactive visual design environments such as Lyra~\cite{Satyanarayan2014}, iVoLVER~\cite{Mendez2016}, and iVisDesigner~\cite{Ren2014}.
Even more recent tools include Data Illustrator~\cite{liu2018dataillustrator} and DataInk~\cite{Xia2018}, both of which use direct manipulation to allow designers to bind visual features to data.
Common among them is that they
require no programming, and are thus positioned at the very other end of the spectrum from visualization toolkits and grammars.
However, interactive view specification can be clumsy and inefficient at times, and may still be plagued by challenges due to the intricacies of visualization design.

Shelf-based visualization environments such as Polaris~\cite{stolte2002polaris}, Tableau~\cite{tableau}, and PoleStar~\cite{polestar} fall in the middle of the spectrum.
Common among these is their ability to allow the user to drag and drop data dimensions, metadata, and measures to specific ``shelves,'' each one representing a visual channel such as axis, shape, scale, color, etc.
This point-and-click approach to visual specification is flexible enough to  construct a wide range of visualizations, but not so complex so as to become technical, such as Lyra, iVoLVER, and iVisDesigner.
In DataSite, we employ a variant of a shelf-based visualization environment for this very reason.

\begin{figure*}
	\centering
 	\includegraphics[width=0.85\textwidth]{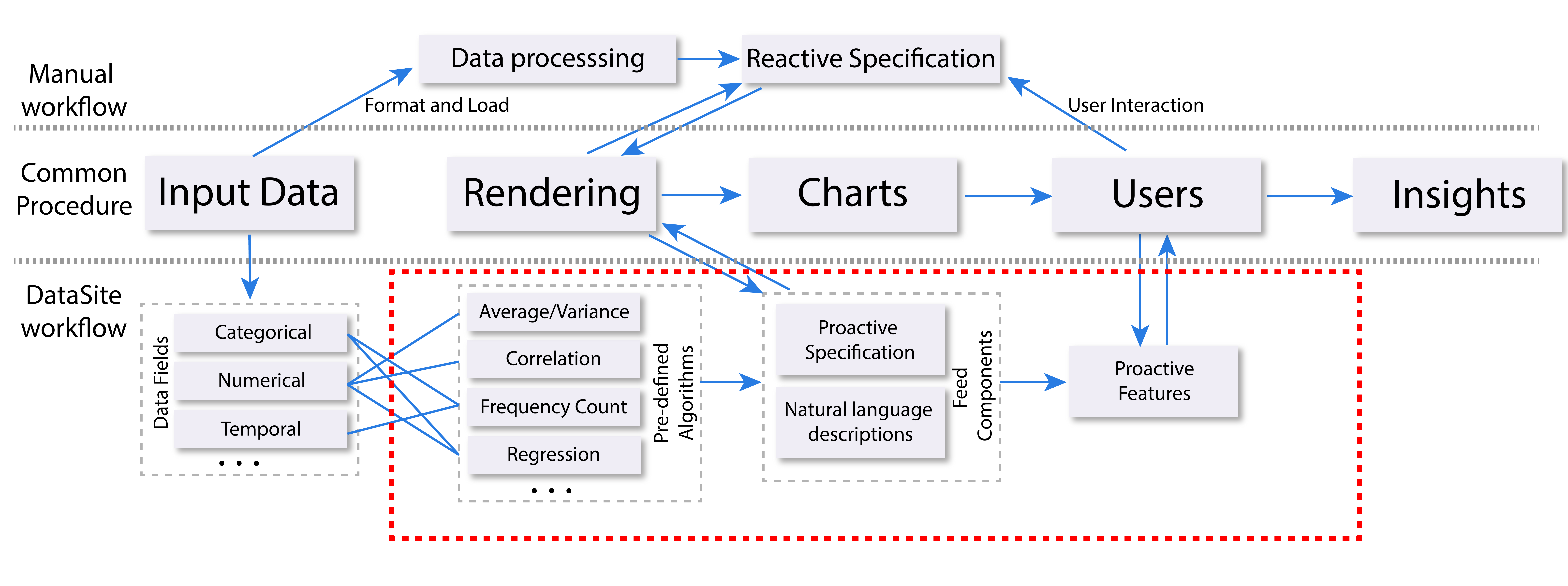}
 	\caption{The structure and workflow of the DataSite visual analysis system.
    The manual and proactive (DataSite) visualization workflow have shared common procedures in the middle.
    Components within the red rectangle are the key parts of DataSite: proactive computational modules that can run through various data fields, visualization, and natural language descriptions.
    These together offer suggested features in the feed.}
	\label{fig:overall_structure}
\end{figure*}

\subsection{Visualization Recommendation}

The idea behind visualization recommendation is to use recommendation engines~\cite{herlocker2004collaborafilter} to suggest relevant views to the user, thus reducing the cognitive load.
While this idea has seen a resurgence in the visualization community in recent years, it is by no means a new idea.
Mackinlay~\cite{mackinlay1986automating} first proposed automatic visualization design based on input data in 1986.
His work combines expressiveness and effectiveness criteria inspired by Bertin~\cite{bertin1983semiology} and Cleveland et al.~\cite{cleveland1984graphical} to recommend suitable visualizations.
Tableau's Show Me system~\cite{mackinlay2007show} provided a practical and commercial implementation of these ideas.

Many similar approaches to automatic visual specification exist.
Sage~\cite{roth1994interactive} extends Mackinlay's work to enhance user-directed design by completing and retrieving partial specifications based on their appearance and data contents.
The rank-by-feature framework~\cite{seo2005rank} sorts scatterplot, boxplots, and histograms in a hierarchical clustering explorer to understand and find important features in multidimensional datasets.
SeeDB~\cite{seeDB2014} generates a wide range of visualizations, and define which ones would be interesting by deviation and scale.
Perry's~\cite{perry2013vizdeck} and Van den Elzen's~\cite{van2013small} work attack the problem that generates multiple visualizations shown with small thumbnails.

Recommendation engines have been used to great effect for visualization in the last few years. 
Voyager~\cite{Wongsuphasawat2016} generates a large number of visualizations and organizes them by relevance on a large, scrolling canvas.
Visualization by demonstration~\cite{Saket2017} lets the user demonstrate incremental changes to a visualization, and then gives recommendations on transformations.
Zenvisage~\cite{siddiqui2016effortless} automatically identifies and recommends interesting visualizations to the user depending on what they are looking for.
Recently, Voyager 2~\cite{Wongsuphasawat2017} builds on Voyager, but supports wildcards in the specification and provides additional partial view suggestions. ``Top-K insights''~\cite{tang2017extracting} provides theory for generating insights, which is the main motivation of our paper.
All of these ideas were formative in our work on DataSite, but our approach takes this a step further by focusing on continuous computation from a library of automatic algorithms, with findings propagated to the user in a dynamically updating feed.

\subsection{Proactive Computation alongside Visualization} 
The idea of proactive visual analytics discussed in our paper builds on the idea to opportunistically run computations in anticipation of user needs, which is observed in Novias~\cite{novais2012proactive}, TreeVersity~\cite{treeversity}, and Analyza~\cite{googleexplore} (\textit{Explore} in Google Sheet).
Novias identifies visual elements of evolving features and provides multiple views in an interactive environment.
TreeVersity provides a list of outliers in textual form, which identifies changes in the data automatically.
The most similar research to DataSite is Analyza, which provides auto-computed features in natural language.
In contrast, DataSite aims to push proactive computation to depth and complexity rather than just simple overall statistics in the dataset.
Furthermore, DataSite pushes features to a feed view that is akin to social media feeds users are already accustomed to.

\section{Design Rationale: Proactive Analytics}
\label{sec:design}

The core philosophy for proactive analytics is that human thinking is expensive, whereas computational resources are (generally) cheap.
Following this philosophy, a proactive approach to visual analytics should automatically run computations in the background and present its features to the analyst in an endeavor to reduce the analyst's cognitive effort during the sensemaking process.
In essence, the solution is to use brute force computational power of the computer to help balance out the equation between the human analyst and the computer tool.
This leverages the respective strengths of each partner while complementing their weaknesses:

\begin{itemize}
\item{Human analyst:} The human operator analyzing data.
  \begin{itemize}[nosep]
  \item\textit{Strength:} creativity, experience, deduction, domain knowledge. 
  \item\textit{Weakness:} limited short-term memory, computational power, lack of analysis expertise, and  limited perception.
  \end{itemize}
\item{Computer analytics tool:} The tool facilitating analysis.
  \begin{itemize}[nosep]
  \item\textit{Strength:} significant memory and computational power; large library of analytical algorithms.
  \item\textit{Weakness:} no creativity, intuition, or deduction. 
  Lack of domain knowledge.
  \end{itemize}
\end{itemize}

Based on these ideas and the related work (see the previous section), we derive the following design guidelines for proactive visual analytics tools (we give examples of illustrative systems for each guideline):

\begin{itemize}
\item[D1]\textbf{Offload computation from analyst to machine.}
The analytical tool should be designed so as to offload as much as possible of the analysis from the user.
Given our core philosophy, this means that the tool should never be idle waiting for the user to act.
Instead, it should always be running tasks in the background, and start another task as soon as one finishes.
\begin{itemize}
\item\textit{Example:} The Voyager~\cite{Wongsuphasawat2016} and Voyager 2~\cite{Wongsuphasawat2017} systems pre-emptively perform computation on features of the current dataset to provide new views to the user.
\end{itemize}

\item[D2]\textbf{Present automated features incrementally with minimal interruption to the analyst.}
Automatic features derived by the background computational processes must be propagated to the user, but the presentation of these features should be designed so as not to interrupt the user's cognitive processes needlessly. 
These features should be accumulated in a feed where they can be easily surveyed and viewed at the user's own initiative rather than in a blocking manner that requires action.
\begin{itemize}
\item\textit{Example:} The InsightsFeed tool~\cite{BadamProgressive} progressively runs calculations in the background and updates the displays as new results come in.
\end{itemize}

\item[D3]\textbf{Reduce the knowledge barrier of human thinking.}
Data analytics is a nascent discipline with rapidly evolving methods, many requiring the data to support specific assumptions or exhibit certain properties, so it is often difficult even for expert-level analysts to stay abreast of current practice~\cite{Batch2018}.
This is another situation where timely proactive support can save analyst effort by investing CPU time: the tool can simply run every conceivable analytical method from a large library of methods (ordered by perceived utility) and only present interesting trends.
\begin{itemize}
\item\textit{Example:} Tang et al.~\cite{tang2017extracting} propose a tool that automatically calculates the top-$k$ insights from a multidimensional dataset based on an importance function used to score different findings.
\end{itemize}

\item[D4]\textbf{Eliminate ``cold-start'' through exposing potentially relevant features of the data early during exploration.}
A challenge related to the knowledge barrier is the so-called ``cold-start problem''; the fact that, when beginning analysis on a new dataset, it can be challenging to know how to get started because the data can be overwhelming and difficult to get a handle on.
Again, this can be mitigated by not choosing but simply performing \textbf{all} applicable analyses from a library of such methods.
\begin{itemize}
\item\textit{Example:} Schein et al.~\cite{coldstartrecommend} define the cold start problem for recommender systems and propose a method for deriving recommendation scores for new items based on similarities to existing items.
\end{itemize}

\end{itemize}

\begin{figure*}
	\centering
 	\includegraphics[width=\linewidth]{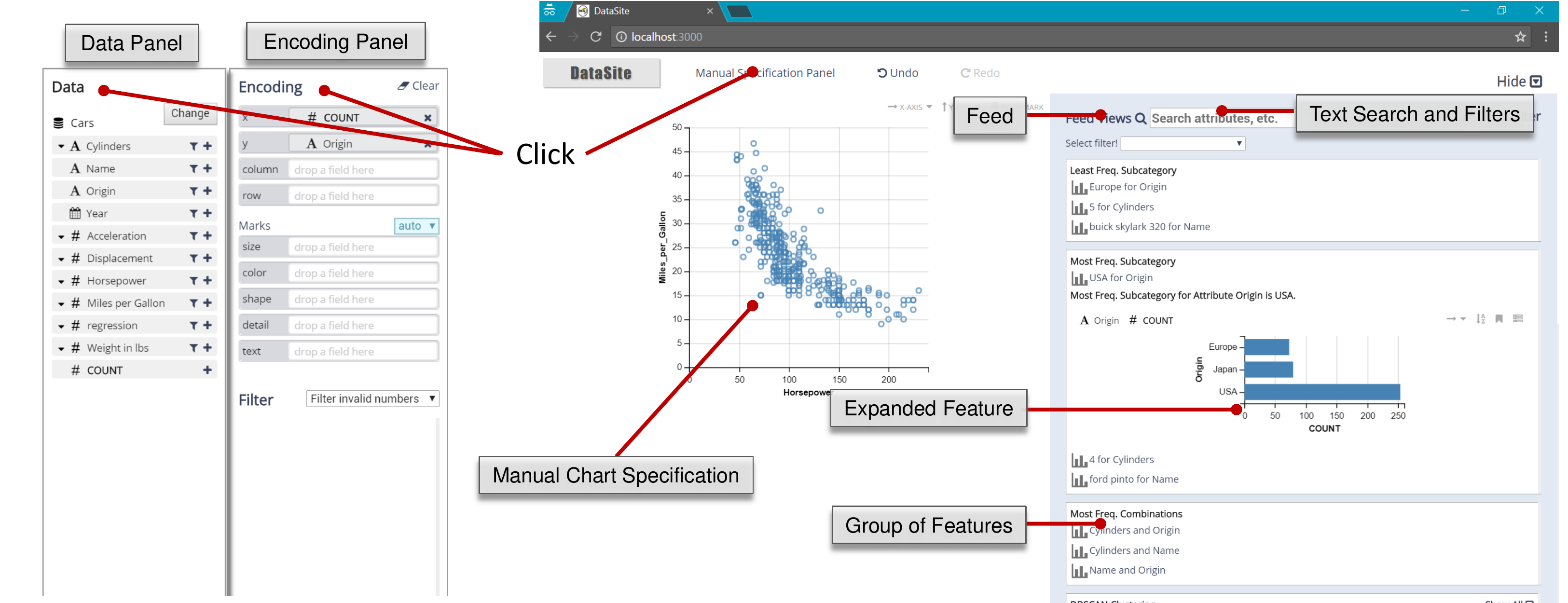}
 	\caption{DataSite is a proactive visual analysis system that allows the analyst to explore data on the web-based client using a standard visualization interface (data, encoding, and manual chart specification panel), while a server-side component automatically selects and executes relevant computations without prompting.
    Features gleaned from these analyses are surfaced and updated dynamically in a Feed View (right) on the client, similar to posts in a social media feed.}
	\label{fig:teaser}
\end{figure*}

\section{The DataSite System}
\label{system}

DataSite consists of (1) a user interface for proactive visual analytics containing components for visualization authoring along with a dynamically updated feed view, and (2) a proactive computation engine continuously running background modules on a target dataset.
The user interface runs in a client on a modern web browser and consists of a manual visualization view coupled with a \textit{feed view}.
The client interface is designed for an analyst to use when manually analyzing data in their web browser.
The computation engine, on the other hand, runs in a server process, thereby offloading computation (D1).
The feed view accumulates features as status updates (D2), each consisting of a title, an icon, a detailed textual description, and a representative interactive visualization.
Working in concert, the feed view reduces the knowledge barrier (D3) by continuously displaying trends from the proactive computation engine.
The feed also provides a starting point, eliminating the cold start problem (D4).

\subsection{Client-Side: Visualization Interface}

The DataSite interface comprises a data schema panel, an encoding panel, a manual chart specification view, and a feed view (Figure~\ref{fig:teaser}).
The data schema, encoding, and chart specification views together compose a basic shelf-based visualization system for exploring the data.
The main visualization view is shown in the center of the screen, with the data schema and encoding on the left.
This interface design is consistent with typical exploratory visualization tools, such as Tableau, QlikView, and Spotfire.

Augmenting this design, the dynamically updated feed view is the key interface-level contribution of DataSite.
The feed accumulates features generated by the server-side computation engine.
To give ample space for the analyst's navigation through the interface components, the feed is placed on the right side of the screen to complement the manual specification view.
The data and encoding panel can be hidden to free up additional space.

\begin{figure}
  \centering
  \includegraphics[width=0.80\linewidth]{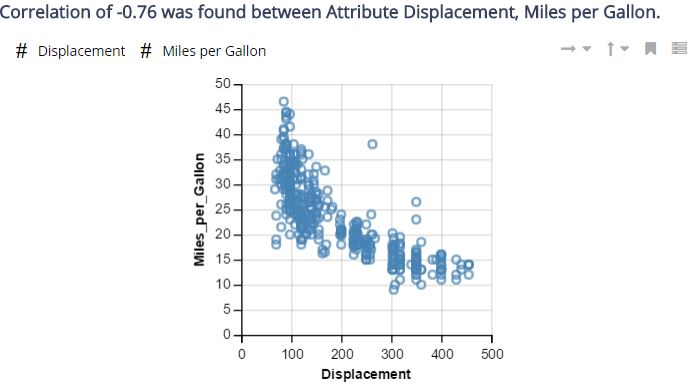}
  \caption{Example of features in the feed: a brief textual description (``Correlation metric between Miles per Gallon and Displacement attributes in a Cars dataset.'') with a corresponding auto-generated chart (scatterplot for these two specific attributes).
  A red line that shows the computed correlation trend between two attributes is also shown.}
  \label{fig:component_example}
\end{figure}

The feed view is inspired by social media feeds, where events {\color{red}pinned} by participants appear in a dynamically updating list in chronological order.
A \textit{data feature} in the feed is a notification from a computation engine.
Once a feature has been computed by a server-side analysis component, it will be dynamically added to the feed.
The feed view can be searched and filtered; sorted by the computational measure, the time it was produced, or in alphabetical order; and grouped by type.
Each feature is initially represented as a short title and an icon explaining the underlying computation task. 
Users can expand a feature to see a detailed text description as well as an associated chart for the data attributes processed by the underlying computation (Figure~\ref{fig:component_example}), and then collapse it when needed.
When the user manually selects or drag-and-drop data attributes in the encoding panel, the feed will be reordered, with computational categories that contain the selected data attributes moved to the top.

\begin{figure}[htb]
	\centering
 	\includegraphics[width=0.3\linewidth]{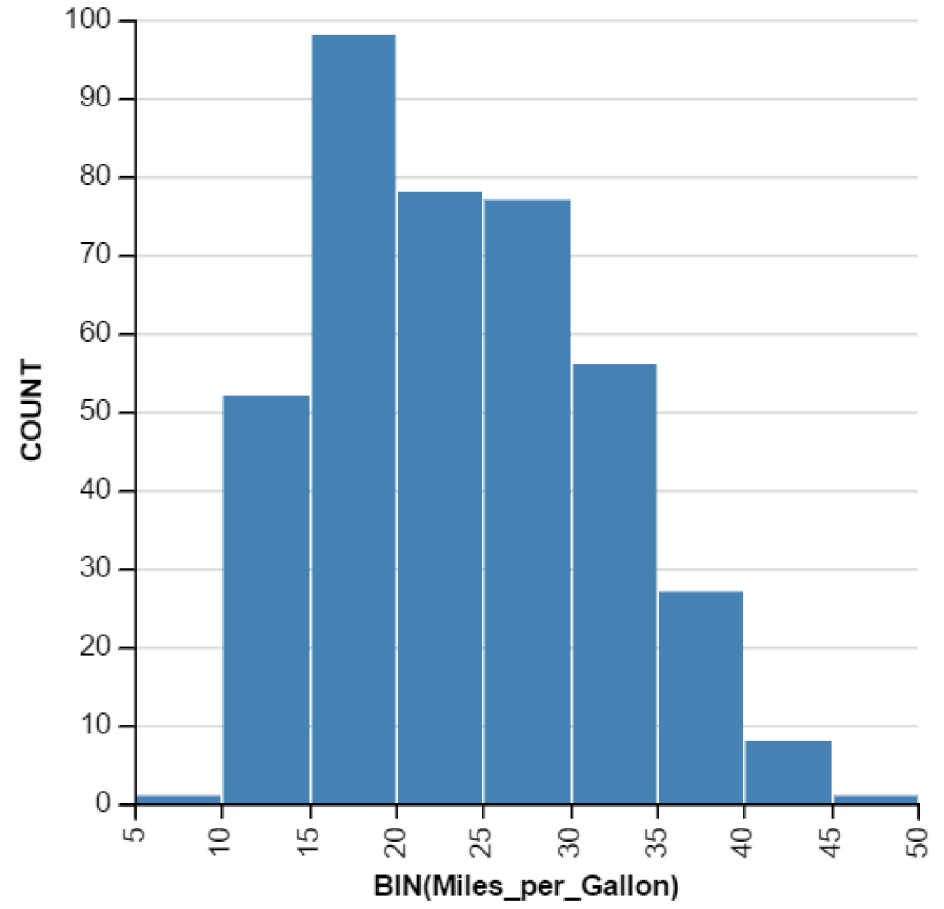}
 	\includegraphics[width=0.3\linewidth]{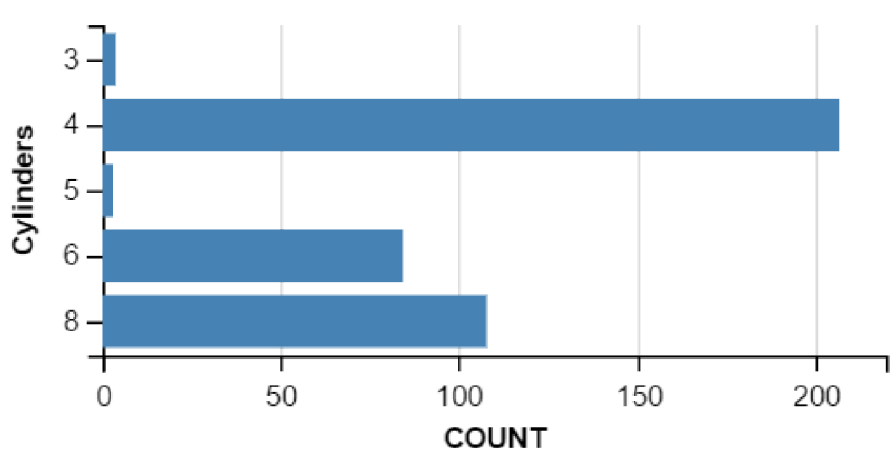}
  	\includegraphics[width=0.3\linewidth]{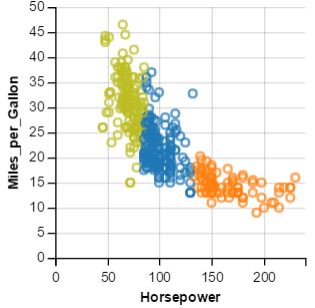}
 	\caption{Chart types for different computational modules used in DataSite.
    From left: histogram bar (mean/variance), histogram line
    (min/max), and scatterplot (clusters in 2D).}
	\label{fig:chart_formats}
\end{figure}

\begin{figure}[h]
	\centering
 	\includegraphics[width=0.80\linewidth]{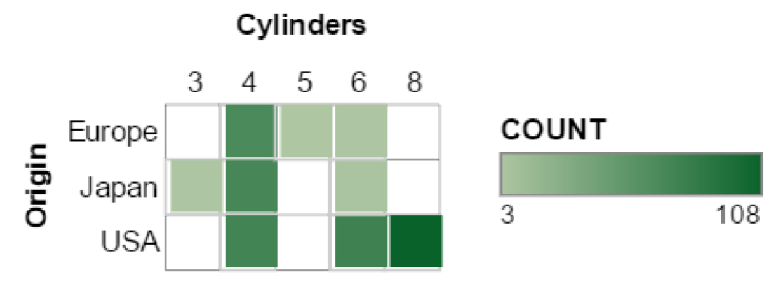}
 	\caption{Representative chart (heatmap) automatically generated for co-occurrence frequency counts of two categorical data fields (origin country and number of cylinders) in a Cars dataset.
    Darker color indicates more counts in that category combination; in this example, V8 cars from the USA.}
 	\label{fig:heatmap}
\end{figure}

Each update item in the feed consists of the following components (expanded on demand):

\begin{itemize}
\item\textbf{Title:} Each update has a compact title that gives a brief idea of the contents of the feature or insight. 
This title and the thumbnail is the only thing shown when the update is collapsed, thus taking a minimum of display space.
For example, the Pearson correlation generates titles such as ``$\rho = 0.5$ for Weight and MPG.''

\item\textbf{Icon/thumbnail:} A small iconic representation that gives a visual indication of the contents. 
For computations that generate charts, this could be a miniature thumbnail of the chart.

\item\textbf{Textual description:} A description of the feature presented on the feed view in a proactive manner.
For example, for the Pearson correlation coefficients~\cite{correlationPearson} between \textit{Weights in lbs} and \textit{Miles per Gallon} in cars dataset~\cite{carsDataset1981Henderson}, the textual description is: ``Correlation of $0.5$ was found between attributes \textit{Weights in lbs} and \textit{Miles per Gallon}.'' 
This active description gives the analyst the sense that the computer is their collaborator in helping them explore the data.
To avoid overloading the feed with an excessive number of features, we combine related trends and illustrate them with a single chart (e.g., min/max are combined, described as a range, and shown on a bar chart, see Fig.~\ref{fig:chart_formats}).

\item\textbf{Charts:} Manual view specification yields full control to the analysts, but may cause high cognitive load.
To avoid this, DataSite shows the most efficient encodings for each chart corresponding to tasks from a computational module according to the existing metrics~\cite{bertin1983semiology, cleveland1984graphical, mackinlay1986automating}.
Charts are lazily rendered when clicked, thus reducing the page load significantly.
For instance, with two categorical attributes, DataSite renders a heatmap (Figure~\ref{fig:heatmap}) with the intersecting frequency counts marked in color.
Similar to the approach in previous research~\cite{badam2016timefork, Wongsuphasawat2016}, charts can be moved to the main view panel by clicking a \textit{specify the chart} icon on the top right.
Furthermore, charts highlight aspects of the underlying computation as visual cues: for example, charts generated from the clustering computation will highlight the clusters within the chart.

\end{itemize}

In addition to automatic updates, analysts can {\color{red}pin} views from the manual chart visualization window, saving that view as an update in the feed.
The feed view keeps track of these user-generated updates as a separate category.
This is the same as bookmarking charts, and in the future we plan to make the feed a collaborative space, where either human or computer {\color{red}pin} features to allow sharing of findings.

\subsection{Server-Side: Computation Engine}
\label{computation_modules}

The server-side DataSite computation engine begins analyzing a dataset as soon as it is uploaded.
The engine consists of multiple computational modules (easily extended as plugins); Table~\ref{tab:algorithms} shows a sample.
A single module can yield several tasks; for example, a simple Pearson correlation module would create a task for each combination of numerical attributes, but not for categorical attributes.

A scheduler analyzes the data and runs computations in a specific order; see the next section for details on scheduling analysis.
The computation engine is multi-threaded using a computational thread pool, executing each computation in the scheduled order.
For each finished task, the computational module will generate a status update that will be pushed to the visualization interface.
As soon as a computational thread is freed up, the scheduler will recycle the thread for a new task.
In this way, the engine is never blocked by complex, long-running tasks.
Furthermore, each computation module executes independently, so a single module failure does not affect the overall system.
For example, if one module fails executing due to errors or invalid data, it will not return results, but other modules can still execute without interruption.

By virtue of this modular architecture, DataSite can be easily extended with new computation modules.
The current implementation provides statistical analysis, K-means clustering (3, 5, 7 clusters), density based clustering (DBSCAN with various parameters), linear regression, and polynomial regression modules.
Figure~\ref{fig:chart_formats} shows sample charts created in the feed view for some computation modules.
Again, computational modules can be added easily, and the goal of the framework is to have as many modules as possible such that computational engine is always running and recommending insights to the user.

\subsection{Scheduling Automatic Analysis}

The scheduler is a core component in the computation engine.
It passes the dataset through its entire library of loaded computational modules, receiving an estimate of the computational complexity and relevance from each module based on the meta-data---number of attributes, types, and dataset size.
Furthermore, the scheduler also encodes typical analytical practice by focusing on main effects and trends in the dataset, and then turn to specific combinations of dimensions of the data.
All of the metrics are then used by the scheduler to determine which modules to run, and in which order to run them.
It may also reschedule jobs in response to results returned from another module; for example, to run post-hoc analysis in response to a significant result from an analysis of variance test.
In addition, the scheduler may choose to launch long-lasting analyses---such as multidimensional scaling or cluster analysis---early, knowing that these results will take a while to return.

DataSite currently utilizes asynchronous multi-threaded operations for all the existing computation modules mentioned above.
The system starts executing all the algorithms asynchronously when initially receiving the dataset.
It then waits for results to come back, updating the feed in response.
In the future, we anticipate letting the user guide the computation order, either more implicitly, or explicitly (by providing interactions to steer the computation).
This would enable customizing the DataSite scheduler to the analytical practice of a specific user while retaining the overall hybrid model.
However, such implicit or explicit computational steering of proactive analysis is outside the scope of our current work.

\begin{table*}
  \centering
  \begin{tabular}{rrclp{7cm}}
  	 \toprule[0.5pt]
    \textbf{Modules} & \textbf{Data Formats} & \textbf{\#Attr.} & \textbf{Chart} & 
    \textbf{Description}\\
     \toprule[0.5pt]
    Mean/variance	& numerical & 1 & hist. (Fig.~\ref{fig:chart_formats}) & Attribute $A$ has mean of $X$ with variance of $Y$. \\ 
    Min/max (range) & numerical & 1 & hist. line & Range (min, max) was found in attribute $A$. \\
    Freq. counts & categorical  & 1 & aggr. (Fig.~\ref{fig:chart_formats}) & $X$ was the most/least frequent sub-category in $A$. \\
    Freq. comb. & categorical & 2 & heatmap (Fig.~\ref{fig:heatmap}) & Most frequent combination was found between $X$ in attribute $A$, and $Y$ in attribute $B$. \\
    Correlation & numerical & 2 & scatterplot & Correlation of $A$ was found between $X$ and $Y$. \\
    $K$-means & numerical & 2 & scatterplot (Fig.~\ref{fig:chart_formats}) & $K$-means with $N$ clusters between $X$ and $Y$ has average error $E$.\\
    DBSCAN & numerical & 2 & scatterplot & DBSCAN between $X$ and Y with minPts = $p$ estimated $K$ clusters. \\
   Linear Regression & numerical & 2 & regression line & Linear regression between $X$ and $Y$ has estimate error of $E$. \\
   Poly. Regression & numerical & 2 & regression line & Polynomial regression between $X$ and $Y$ has estimate error of $E$. \\
     \toprule[0.5pt]
  \end{tabular}
  \caption{Example computational modules with corresponding data and chart types.
  We have currently used algorithms working with one or two data attributes
  in our computation engine. 
  Brief textual descriptions for each module are also listed.}
  \label{tab:algorithms}
\end{table*}

\subsection{Implementation}
\label{sec:implementation}

DataSite is based on a client/server architecture.
The client side is developed using AngularJS,\footnote{\url{https://angularjs.org/}} a JavaScript-based web application framework.
The visualization functionality in the DataSite client is based on the PoleStar interface (available as open source)~\cite{polestar}, which is built on top of Vega-Lite~\cite{Satyanarayan2016_vega}.

We implemented the computational engine using Node.js,\footnote{\url{https://nodejs.org/}} a non-blocking server-side JavaScript framework.
Datasets of interest can be uploaded by the user on the client interface, and sent to the server.
The server processes them using the engine and proactively sends the finished features to the feed view.
This structure enables managing a wide array of input data formats, and scales to large datasets. 
In essence, the server does all the heavy lifting: loading data, maintaining the connections to clients, executing computational modules, and updating features.

\begin{figure}[htbp]
	\centering
 	\includegraphics[width=0.90\linewidth]{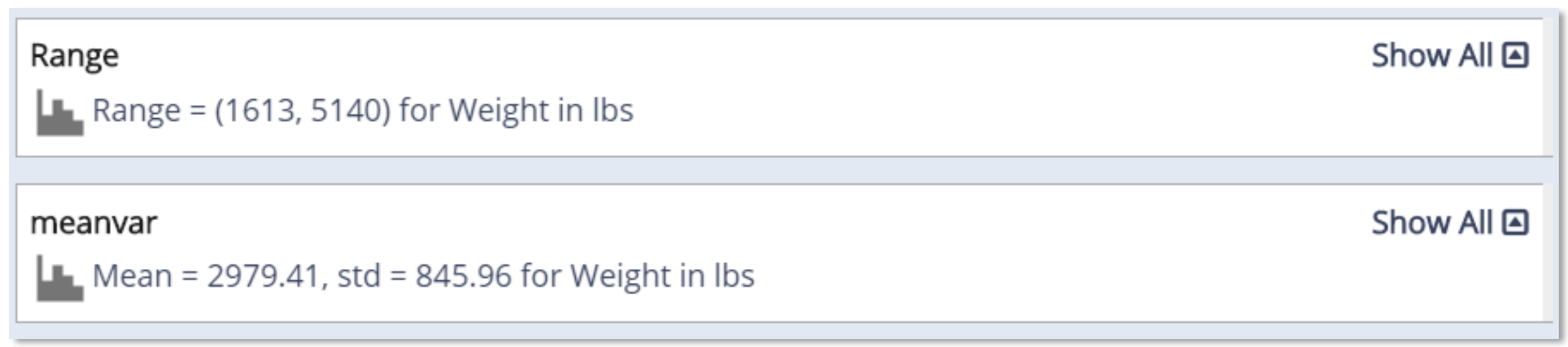}
 	\caption{Feed items for average weight in lbs, as well as range (in lbs).}
 	\label{fig:feeditems_rangemean}
\end{figure}

\section{Usage Scenario}
\label{sec:usagescenario}

Here we illustrate how an analyst can use the DataSite system to examine and find interesting features and insights about the class car dataset~\cite{carsDataset1981Henderson}.
As soon as the analyst uploads the dataset into DataSite, the system will queue up computations scheduled by their suitability for the specific dataset.
Meanwhile, the client-side shows the interactive interface to enable the user to begin data analysis (Fig.~\ref{fig:teaser}).

From the data panel (Fig.~\ref{fig:teaser}, left), the analyst will see the dataset has three categorical attributes, one temporal, and six numerical attributes.
To encode the field, the analyst can manually drag-and-drop or auto-add it to the encoding panel (Fig.~\ref{fig:teaser} middle) to create a visualization. 
To start exploration, he/she may want to get an overview of the dataset.
Without creating any visualization manually, DataSite updates with notifications that are automatically computed by the server in the Feed field on the right.
The analyst finds that the average weight of cars is 2,979 lbs, and the range is from 1,613 to 5,140 lbs (Figure.~\ref{fig:feeditems_rangemean}).
The feed is dynamically updated with notifications once each computation is finished.
There are simple statistics for numerical attributes, such as mean and variance, ranges (which are min and max), and correlations, as well as frequency counts for categorical attributes.
Notification in the feed include a brief natural language description.
The analyst may be interested in why the \textit{Displacement} and \textit{Miles per Gallon} has a correlation of -0.76.
The analyst clicks on the notification, causing it to be expanded to show a corresponding chart (Fig.~\ref{fig:component_example}) explaining the finding: As the \textit{Displacement} increases, \textit{Miles per Gallon} decreases.
In order to view details and conduct further modifications to the chart, the analyst can move the chart into the main view (Fig.~\ref{fig:teaser} right) by clicking the icon on the top right.

The analyst may also want to understand the highest frequency counts for the number of cylinders.
To achieve this, the analyst uses the filter drop-down option, clicks on the Frequency Count filter, and types ``cylinders'' into the text bar.
The feed will filter the precise results: cars with 4 cylinder engines has the highest frequency.
The analyst can then pin interesting manual specification charts in the feed, which can be used for tracking the progress of analysis and re-visiting the analysis in the future.

\section{Evaluation Overview}
\label{sec:evaluation_overview}

DataSite creates a new method for visual exploration through a mixture of manual and automated visualization specifications driven by proactive computations. 
For this reason, we are interested in understanding whether the exploratory analysis with DataSite supports bootstrap understanding and broad coverage of the data. 
We are also curious about knowing how/why the feed helps, and how it changes the analyst's approach in finding features. 
To answer these questions, we conducted two user studies: (1) comparing with a manual visualization specification tool, PoleStar, focusing on data field coverage; and (2) comparing with a visualization recommendation system, Voyager 2~\cite{Wongsuphasawat2017}, focusing on data exploration to compare the effects of adding a \textit{Feed} (in DataSite) versus \textit{Related Views} (in Voyager 2).
In other words, Study 1 aims to understand the fundamental utility of the feed view itself, while Study 2 expands this to understanding DataSite's proactive analytics workflow compared to a recent visual recommendation system.

\subsection{Dataset}

To enable comparisons of our results with PoleStar and Voyager 2, we reused the same datasets for our studies. 
One is a collection of films (``movies") containing 3,178 records and 15 data fields, including 7 categorical, 1 temporal, and 8 quantitative attributes. 
The other dataset contains records of FAA wildlife airplane strikes (``birdstrikes''), which contains 10,000 records and 14 data fields, with 9 categorical, 1 temporal, and 4 quantitative attributes. 
These two datasets have similar complexity (w.r.t.\ number of attributes), and are easy to understand. 

\subsection{Study Design and Procedure}

In both user studies, we used $2$ tools with $2$ datasets (one dataset on each tool interface). 
Participants in both studies started with an assigned tool and dataset, and then moved to the second interface.
To deal with learning effects, we counterbalanced the order of tools and datasets---half of our subjects used PoleStar/Voyager 2 first and the other half used DataSite first (similarly with the dataset).

Each participant began a session by completing a short demographic survey.
However, we did not screen participants based on the demographic information provided.
The participant was then introduced to the first interface assigned. 
The participant were first shown the interface and a tutorial on how to use the tool with the classic automobile dataset for training purposes.
For DataSite, they were also shown the feed view and its associated operations. 
The participant was then allowed to train using the interface with the automobile dataset, and were encouraged to ask questions about the dataset and tools until they indicated that they were ready to proceed.

The experimenter then briefly introduced the participant to the experimental dataset and asked him/her to explore the dataset ``as much as possible'' (open-ended) within a given time of 20 minutes for each system.
They were asked to speak out aloud their thinking process and insights. 
We did not ask the participants to have specific questions to answer during the session, as this may bias them in exploration and limit their focus to specific subsets rather than the whole dataset.
After completing a session with the first tool, the participants repeated the same procedure for the second tool and dataset.
After completing the tasks for both tools, they were asked to complete a questionnaire with Likert-scale ratings on the efficiency and usefulness of each tool as well as the participant's rationale for their ratings. 
Participants were also encouraged to verbalize their motivations and comments on each tool.
Each session with two studies as well as exit survey in total lasted around 60 minutes.

All the sessions were held in a laboratory setting in a university campus. 
Both tools ran on Google Chrome web browser on a Windows 10 laptop with a $14$-inch display. 
The experimenter observed each session and took notes. 
Participant's interactions with the tool were logged into files, including  application events. 
The audio of the session was also recorded for further analysis.

\section{User Study 1: Comparison with PoleStar}
\label{sec:study1}

In this study, we compare DataSite with a Tableau-style visual analysis tool (PoleStar). 
As described earlier, this study was motivated by a fundamental question: what happens when you incorporate a feed view into a conventional visualization tool. 
We therefore studied the data field coverage during open-ended visual exploration influenced by the Feed in DataSite against Polestar (a baseline interface without the Feed view).
Note that apart from the Feed view, the DataSite interface resembles the PoleStar interface.
Our hypotheses were: (1) DataSite would have higher data field coverage and more charts viewed, (2) DataSite would allow exploration of complex charts with multiple encodings (capturing multiple attributes), and support faster understanding of the data.

\subsection{Participants}

We recruited 16 paid participants (7 female, 9 male) from the general student population at our university.
Participants were 18 to 35 years of age, with some prior data analysis and visualization experience.
All of them had experience with data analysis and visualization tools: All (16) had used Excel, 10 had used Tableau, 7 Python/matplotlib, 7 R/ggplot, and 3 had used other analytics tools. 
No participant had previously seen or analyzed the datasets used in our study. 
They had not heard of or used DataSite or PoleStar, though some found PoleStar to be similar to Tableau. 

\subsection{Results and Observations}

As mentioned in the evaluation overview, participants' interaction logs and notes taken by experimenter were collected during the study.
We used the linear mixed-effects model~\cite{barr2013random, green2013efficacy} for our analysis of the collected data. 
We modeled the participants and datasets as random effects with intercept terms (per-dataset and per-participant bias), and regarded different tools and the order of tool usage as fixed effects. 
This setting accounts for the variance of tools and datasets with individual subject's performance during the study.
We used likelihood-ratio tests to compare the full model with other models to evaluate the significance of difference.

Overall, correctness in collected insights was high across both conditions (PoleStar and DataSite), with no reliable difference.
For this reason, we chose to disregard further analysis of this effect.

To assess the broad coverage of data fields, we consider the number of unique data field sets. 
Users may have been exposed to a large number of visualization charts, 
while the unique field sets shown and interacted with are conservative and reasonable measures of overall dataset coverage. 
Based on this, there is a significant improvement of data attribute coverage with DataSite (30\% increase compared to PoleStar: $\chi^2(1) = 19.26$, $p < 0.005$).
Participants interacted with more charts, both from the feed as well as by modifying encodings from the charts present within the feed.  
This confirms the first hypothesis.


There are more multi-attribute charts (encoding two or more data attributes) that participants viewed and interacted with using DataSite than PoleStar ($\chi^2(1) = 10.31$, $p < 0.005$).
This is expected since DataSite provides pre-computed features, while participants had to manually create all visualization charts themselves in PoleStar. 
75\% participants have seen at least 50\% more data fields in DataSite.
Participants also found twice the number of charts using DataSite that are informative and worth ``speaking out'' ($\chi^2(1) = 7.82$, $p < 0.005$).
10 participants have created more than 3 advanced charts with the help of feed (and ``spoke out'' about them): they started with charts from feed and added more data fields as encodings to the charts. 
This suggests that the DataSite system through its Feed view leads to the users viewing more number of charts that are beneficial from their perspective. 
It also indicates that DataSite encourages the user to reach complex (multi-attribute) charts during visual exploration, confirming our second hypothesis.

Participants showed great interest in the features within the feed view. 
Most of them spent at least 25\% of time on exploring the feed itself. 
All participants felt that the feed is useful for analysis and provides guidance of ``where to look'' in the data. 
They rated DataSite higher than PoleStar in terms of efficiency (Likert scale, 1 to 5, mean: 4.67 vs 3.40) and comprehensiveness (mean: 4.20 vs.\ 3.21).
All participants rated the usefulness of the feed 3 or higher.

\section{User Study 2: Comparison with Voyager 2}
\label{sec:study2}

The results from the first study were promising and they answer our fundamental questions about the utility of the DataSite feed view.
In Study 2, we compared DataSite with Voyager 2, a modern visualization recommendation system.
The goal was to observe differences and further understand the utility of the feed in DataSite compared to the Related Views and \textit{wildcards} in Voyager 2. Our hypotheses are: (1) DataSite will provide comparable if not more data field coverage owing to its rigorous computation engine; and (2) DataSite will better guide the user's exploration towards faster and comprehensive understanding in the given time.

\subsection{Participants}

We recruited 12 participants (8 female) from our university.
All had similar demographics (between 18 and 35 years of age) and data analysis experience as before: all participants (12) had used Excel, 8 Tableau, 6 Python/matplotlib, 1 with R/ggplot. 
They had not heard of DataSite or Voyager 2, or seen the datasets involved.

\subsection{Results: Quantitative}

We used the same linear mixed-effect model for statistical analysis in Study 2 similar to Study 1.
As for Study 1, participant insights were collectively accurate independent of condition (DataSite and Voyager 2); thus, we chose not to analyze this aspect further.

\subsubsection{Data Field Coverage}

\begin{figure}
	\centering
 	\includegraphics[width=0.70\linewidth]{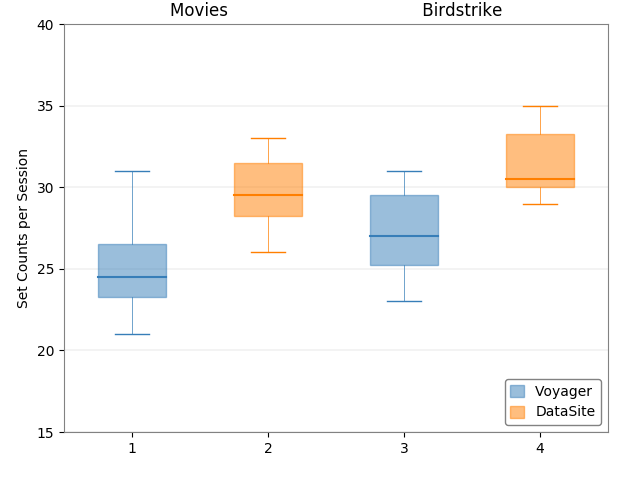}
 	\caption{Box plot showing the distributions of unique fields that users interacted with per tool and dataset.
    DataSite has slightly larger number of unique field sets in both cases.}
	\label{fig:boxplot}
\end{figure}

We first looked into the participants' performance separately for both datasets (movies and birdstrikes), and compared the effects of visualization tools.
We consider the number of unique field sets that users have shown and examined, respectively (similar to the previous study). 
In Figure~\ref{fig:boxplot}, we see that for movies and birdstrikes datasets, the number of unique field sets that users interacted with (hovered mouse for more than three seconds) is similar: DataSite has 5 and 4 more unique field sets respectively in the birdstrike dataset (median: 30 in DataSite vs.\ 25 in Voyager 2) and movies dataset (median: 31 in DataSite vs.\ 27 in Voyager 2). 
Overall, DataSite promotes slightly more data field coverage in total (mean: 30 and 26), mainly because the feed contains an exhaustive list of features across computational modules. 

In regard to the number of unique field sets that have been shown (the user may look through the charts without interaction) to the users, DataSite users (mean 43, s.d.~19.7) were shown fewer charts than Voyager 2 (mean 54, s.d.~13.5). 
The reason may be that Voyager 2 shows charts by default, while DataSite needs user interaction to expand the features in the feed to see the charts. 
As for the number of charts that participants spoke out aloud during the study, the tools have a significant difference ($\chi^2(1) = 7.34$, $p < 0.05$): DataSite (mean 14.53, s.d.~2.04) gave participants 30\% more charts to ``speak out'' about, compared to Voyager 2 (mean 11.63, s.d.~2.32). 
In other words, participants found more charts to be informative and worth talking about using DataSite. 
Among all the ``speak out'' charts, an average of 35\% are directly from the feed.
Other ``speak out'' charts in DataSite are either moved from the feed to the main view and then edited, or manually created.
This indicates that the feed view contributes to more data field coverage and more charts that analysts find useful and worth pointing out. 

When using DataSite, all participants viewed and interacted with charts in the feed. 
Most of them (11 of 12) spent more than 30\% percent of time exploring the feed.
Two participants even used the feed as the main interface for exploring the datasets. 
Beyond this, two participants interacted with more than 70\% of total charts, and 75\% of  their ``speak out'' charts were directly from the feed. 

\subsubsection{Text Search and Filter Usage}

We analyzed the usage of filters and text search bar.
We were interested in observing whether filters and text search can aid them in searching for desired features within the feed view, and whether it is efficient and easy to use compared to \textit{Related Views} and \textit{Wildcards} in Voyager 2.
All participants have used the drop-down filters at least 5 times, and 9 of 12 tried text search. 
8 of 12 of them said that the filters and the text search were useful for quick search of the feed during the study session. 
7 of 12 had used the combinations of text search and filter. 
Three participants found \textit{wildcards} in Voyager 2 to be not very intuitive.
They used wildcards fewer times during the exploration, which matches the results from Wongsuphasawat et al.~\cite{Wongsuphasawat2017}.
In comparison, filters and search options not only contribute to fast data exploration, but also improve the efficiency of drilling down into features during proactive visual analytics. 
This is one of the advantages of providing descriptions for features in the feed view.

\subsubsection{User Ratings}

We collected user's feedback and ratings for tools in the post-study survey. 
For each tool, participants were asked to evaluate the tools based on the efficiency, enjoyability, and ease of use, on Likert scale ratings from 1 (least) to 5 (most). 
The participants rated DataSite ($\mu = 4.32$, $\sigma = 0.67$, $p = 0.14$) higher than Voyager 2 ($\mu = 3.92$, $\sigma = 0.67$) regarding the efficiency.
For enjoyability and ease of use, the ratings are comparable: enjoyability (DataSite: $\mu = 4.33$, $\sigma = 0.65$; Voyager 2: $\mu = 4.08$, $\sigma = 0.67$), ease of use (DataSite: $\mu = 3.92$, $\sigma = 0.85$; Voyager 2: $\mu = 4$, $\sigma = 0.60$). 
When asked about the comprehensiveness of their explorations of the dataset (DataSite: $\mu = 4.42$, $\sigma = 0.87$; Voyager 2: $\mu = 3.75$, $\sigma = 0.51$, $p = 0.013$), $7/12$ users rated DataSite higher and $4/12$ rated both tools with the highest ($5$) score. 
Two participants gave lower ratings for DataSite compared to Voyager 2 and mentioned that it is because they felt in Voyager 2 it was easier to browse multiple charts while in DataSite they had to explicitly click.
Overall, DataSite was seen to be more efficient and presenting a more comprehensive coverage of the data fields with respect to visual exploration than Voyager 2, while maintaining the similar level of enjoyability and ease of use.

\begin{figure}[htb]
	\centering
	\includegraphics[width=0.80\linewidth]{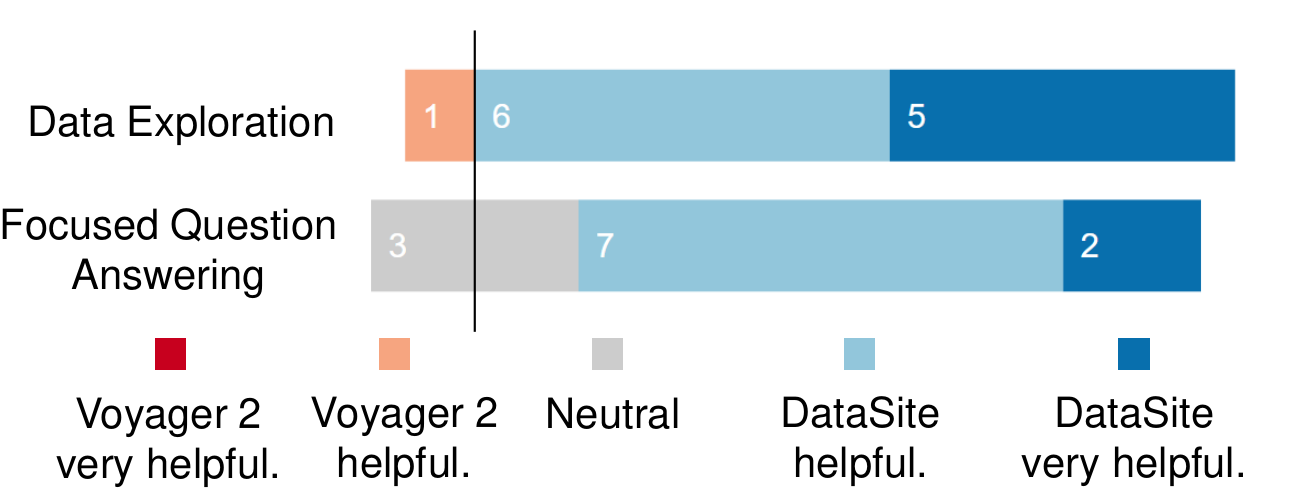}
 	\caption{User preference in terms of visualization tools for open-ended exploration and focused question answering. 
 	DataSite received higher preference in both; 11 of 12 participants prefer DataSite for data exploration, and 9 of 12 prefer DataSite for focused question answering.}
	\label{fig:boxplot_rating}
\end{figure}

Users also responded very positively when asked whether features in the feed provide guidance in their data analysis: 50\% chose 5 and the rest chose a 4 rating.
When it comes to comparison (Fig.~\ref{fig:boxplot_rating}) between two tools on a 5-level symmetric scale (with range $[-2, 2]$.), most participants (11 of 12) preferred DataSite ($\mu = 1.25$, $\sigma = 0.87$) to be most useful or useful for data exploration. 
Beyond this, participants were asked about their preferences between the two tools for focused question answering (as questioned by Wongsuphasawat et al.~\cite{Wongsuphasawat2017}).
7 of 12 users preferred DataSite, and 4 were neutral with no preference, with 1 preferring Voyager 2 (rated -1). 
This is a little surprising since DataSite was primarily designed for visual exploration (and not question answering).

\subsection{Results: Qualitative}

To better understand the results from the statistical analysis, the participant ratings, and how DataSite helped participants explore the datasets, we present our observations below.

\subsubsection{Comparing Charts and Features}

DataSite and Voyager 2 are qualitatively different in that the Voyager 2 recommendation engine stems from query generation and partial view specification using wildcard specifiers~\cite{Wongsuphasawat2017}, whereas DataSite is based features computed by specific computational processes (Table~\ref{tab:feed}). 
Thus, the primary output of Voyager 2 is a sequence of charts generated from this query engine, each arranged in a scrolling ``related views'' panel that are derived from the specified view.
DataSite, on the other hand, generates notifications in the feed view based on results returned from the computational modules loaded.
For this reason, it is difficult to directly compare features in the DataSite feed with chart recommendations in the Voyager 2 interface.
Nevertheless, Voyager 2 is the closest baseline we have, so even if the comparison is not entirely apples to apples, we think that the general metrics of data coverage, insights, subjective rankings, and qualitative feedback that the Voyager 2 evaluation~\cite{Wongsuphasawat2017} uses are appropriate.

Nevertheless, there are many common charts in both DataSite and Voyager 2.
One simple example is that in DataSite, there are descriptive analyses that shows the frequency counts of categorical attributes.
This is also shown in Univariate Summaries in Voyager 2.
When specifying two ``Quantitative Fields'' as \textit{wildcards} in Voyager 2, the specified views contain the scatterplot combining two numerical fields.
If there is a correlation between the fields, this is easily seen using visual inspection of the scatterplot.
For DataSite, this is shown with a description of the correlation and a trending line of regression estimate in a scatterplot, which helps users understand the data.
This is an example of how the query generation engine in Voyager 2 can yield similar results as a directed analytical component in DataSite.
We explore more about the pros and cons of this difference in the discussion.

\subsubsection{Comparing User Findings}

As mentioned above, DataSite features are based on computational algorithms, and it has more ``advanced and detailed'' analysis (quoted from a participant) than standard visualization tools.
Several participants showed great interest in the regression estimate and clustering visualization.
One participant said that clustering chart gave him the clear indication that most birdstrike accidents had very small damage and only a few caused severe outcomes.
Based on this insight highlighted by DataSite, the participant was easily able to dig deeper into which accidents yields the most serious damage.

As an example for the regression line, participants learned its use based on how a car's horse power increases with the number of cylinders.
DataSite creates visualization from algorithmic and analytical perspectives, which may be closer to human thinking, compared to Voyager 2, that generates charts from data attributes using a query engine. 
In more general terms, DataSite provides a suite of computational components that uncover the underlying relationship within the dataset, which may not be easily seen if using manual view specification.
Participants mentioned that the feed provides ``much more detail'', while Voyager is ``basic'' and ``does not give a log of insights'', and sometimes that charts in Voyager 2 ``do not make much sense''.

On the other hand, DataSite features are intrinsically limited by the computational components currently available and loaded in the system.
This is to contrast with the chart-generating feature of Voyager 2, where the power of visualization can yield answers to questions not provided by specific modules.
We go deeper into this discussion at the end of this paper.

\subsubsection{When Participants Used the Feed}

The 12 participants were divided evenly to have different orders of the tools (DataSite first or Voyager 2 first).
Four out of six who used Voyager 2 first, examined the feed (first interacted with the feed) in the beginning of their analysis with DataSite. 
For those exposed to DataSite first, 5 of 6 did the same. 
The rest started their manipulation first with manual specifications. 
It is worth noting that when the participants did not have any idea of how to construct interesting charts to get insights, they (8 of 12) switched to the feed for charts and inspirations (during the middle 10 minutes). 
10 of 12 scanned through the feed at least once in the last 5 minutes of the session. 
9 of 12 participants returned to the feed at least 3 times during the study.
All of them specified at least 3 charts from the feed into the main view.  
This suggests that the feed can help analyst in multiple phases of exploration. 

\subsubsection{In-depth Data Exploration}

Users usually create charts in manual specification tools with less than three attributes for encodings to limit the information encoded to a perceivable level. 
7 of 12 participants found more advanced charts (3 or more data fields/attributes, the same below) that they ``spoke out'' in DataSite than Voyager 2 (at least 20\% more). 
They mentioned that the summary in feed provides descriptive analysis, while charts alone in Voyager 2 may need more time to understand.
It is worth noting that one participant used feed as the only interface for data exploration without additional manual specifications, and none did the same in Voyager 2. 
She explained that the feed provides a systematic approach towards analyzing the dataset, while she had difficulty understanding \textit{Related Views} in Voyager 2. 

\begin{table}[htb]
	\centering
  	\begin{tabular}{cccccc}
        \toprule
        \textbf{\#Charts} & \textbf{Simple stats}
        & \textbf{Corr} & \textbf{Freq} & \textbf{Clust} & \textbf{Regr}\\
        \midrule
        mean	& 2.25 & 4.38 & 4.31 & 3.54 & 3.26  \\ 
        std. dev. &1.25 & 2.5 & 3.46 & 1.02 & 1.57 \\
        \bottomrule
	\end{tabular}
  	\caption{Mean and standard deviation of participant interactions with computational results.
    Participants interacted with advanced features more (e.g., correlations, frequency counts, clustering, etc), while few features regarding simple statistics (min/max and mean/variance) were examined.}
	\label{tab:feed}
\end{table}

\subsubsection{``Speaking Out'' Charts in the Feed}

The number of ``speak out'' charts that users verbally referred to during the study revealed interesting aspects for data analysis by general users. 
Table~\ref{tab:feed} gives mean and variance of features in different categories that the participants ``speak out'' about. 
Participants were more interested in plots of multiple numerical fields and categorical fields, rather than a single numerical field. 
Specifically, they merely viewed the charts in range/mean/variance modules (average number of charts are around 1), and from our observations, they skimmed through the natural language descriptions but did not click to see the charts. 
This implies that simple statistics are not interesting enough for analysts to examine, or the text descriptions alone are sufficient to understand. 

For complex computations, charts are viewed more by expanding their textual description in the feed.
This is because there are usually no intuitive attribute combinations to creating informative charts with data fields (participants had to rely on random combinations or based on their general understanding). 
After seeing the charts in the feed, they all agreed that those charts were more informative than the ones they created by manual view specification.
This is a good confirmation of the utility of proactive visual exploration.

\subsubsection{Inspirations from the Feed}

The feed view provides recommendations for visual data exploration from an analytical perspective. 
The features suggest certain combinations that yield effective visualizations.
All the participants manually specified similar charts (w.r.t. encodings) after they had seen the charts within the feed, especially heatmaps representing frequency combination of two categorical fields. 
More than 80\% (10 of 12) of the participants mentioned that the feed gave them some ideas of which features and encodings can be used to make the chart more informative. 
On the other hand, \textit{Related Views} in Voyager 2 show visualization recommendations to users that can be easily browsed, but participants thought of them just as related charts rather than specific analytical insights.
They browsed through \textit{Related Views} a lot but had never considered about how and why the specific chart was suggested. 
Also, 2 participants felt that the descriptions sometimes were not very easy to understand.

\subsection{Participant Feedback} 

In this section, we list comments, suggestions, and feedback from the free text comments in the post-study survey and audio recording transcripts.
For example, participants described that DataSite helped visual data exploration process:
\textit{``The feed helps gear you in the right direction, especially if you are new to a dataset.
It tells you something notable that is worth looking into.''}
As for comparisons to Voyager 2, \textit{``DataSite is more specific because it gives you the options with various kinds of results.
The feed is very helpful in data analysis.''}
One participant even remarked that \textit{``[DataSite] will be very useful for day-to-day usage, especially for advanced data analysis, and can be used in industrial applications.''}






Overall, the feed view was lauded, with one participant noting that \textit{``the feed in DataSite provides a good starting point to visualize data if you don't have any idea about the dataset.''}
However, participants also provided suggestions on how to improve the feed.
Said one participant, \textit{``it would be better to make feed more user friendly, such as drag-and-drop to move charts into the main view.''}
The feed was also perceived to be daunting, or as one participant put it: \textit{``the feed is very useful, but sometimes it has a lot of results and can be a little overwhelming.''}
Another participant said that \textit{``in DataSite it is a bit difficult for me to understand the results in the feed, while Voyager 2 provides intuitive charts.''}
One participant suggested that \textit{``it would be interesting if there were guided tips that can help when I'm stuck in a chart, such as `try changing x and y axis' when the axis label is difficult to read.''}

\section{Discussion}
\label{discussion}

Our results show that the feed interface in DataSite expedites the process of data exploration both in breadth and depth: participant preference for both open-ended as well as focused exploration was favorable to our tool.
Below we explain these results in depth, and then discuss some of the limitations of our work.

\subsection{Explaining the Results}

Compared with the study results in Voyager 2, DataSite has a comparable unique field set coverage.
The reason why DataSite does not improve the coverage significantly is that Voyager 2 shows all the charts by default, while DataSite only shows charts on demand when participants click on the descriptions.
In other words, DataSite requires participants to actively examine the charts in the feed rather than merely browsing them in Voyager 2's \textit{Related Views}.
Most participants preferred DataSite for data exploration, and rated the feed very useful to aid data analysis and provide trends and guidance of creating meaningful visualization.
It is worth noting that DataSite also yielded higher ratings in focused exploration.
While DataSite is not designed primarily for targeted exploration, the study reveals a potential effect on that.
This also motivates us to consider what and how a targeted data analysis system should adjust, and what evaluations can be done to achieve that purpose.

One observation from our evaluation studies is that simple statistics (average, range, variance, etc) did not interest participants much.
A comprehensive evaluation of what features would be more interesting to the analysts is needed.
The salient features lower barrier for bootstrapping exploration.
However, too many features may distract user's interest, which have to be balanced carefully.
While Voyager 2 also provides efficient visualization recommendations, results from our evaluation indicate that participants felt that the feed was more targeted and worth analyzing.
Three participants noted that while they were going through Voyager's related views, they sometimes forgot what they had seen using manual view specifications.
We speculate that DataSite explicitly labels the features using a textual description facilitates more targeted analysis.

It is worth noting that DataSite exhaustively applies computations to all the possible data fields (and combinations).
While this enhances data coverage, not all modules and corresponding charts represent a clear insight.
For example, categorical attributes such as ``name'' may have thousands of entries, and it is very difficult to find salient trends via such a chart.
While DataSite ranks features by their significance, a more precise saliency measure is needed.
The challenge is how to measure the efficiency of analytical features from a human perspective, and how to unify the metrics across various types of computations.
This requires comprehensively measuring the efficiency for each visualization.
This is further complicated by the fact that different analysts may have different perspectives, or the same analyst may have different perspectives depending on the question in the study.
For the automobile dataset, buyers may wish to see which car is more economic and safer (higher fuel mileage and fewer accident records), while sellers may be interested in popularity (higher profits and larger number of sales).
These contexts should also be considered for personalizing features.
Automatic guided tooltips, suggested by one participant, would be one way to achieve this.

\subsection{Limitations}

Our goal with DataSite is to take computational guidance to its logical extreme, building on the current trend of recommendation engines for visualization.
However, this kind of automatic analysis approach is fraught with challenges, including eroding an analysts' independent thought process (as discussed by Wongsuphasawat et al.~\cite{Wongsuphasawat2017}), automating key decisions that would benefit from analyst insight, and even HARKing~\cite{Kerr1998} (hypothesizing after results are known) and p-hacking~\cite{Selvin1966} (extensively mining a dataset in search of significant findings).
We do not claim that DataSite's mixed-initiative method is optimal for balancing the analytic burden between analyst and computer, only that it is one instance in the design space that shows promise.

However, while DataSite automates some of the analytical process, it does not aim to replace the analyst.
Data analysis is best performed with an analyst in the loop, and DataSite ensures the analyst is always in control.
From our evaluation, the average number of insights from different data sources are 8 manually created, 11 from the feed, and 7 from the feed with modifications.
This observation shows that participants generated almost the same number of insights from feed (automatic) and manually created.

Another valid point of criticism is that computational power is not always cheap; some algorithms are simply not tractable to be run for an entire dataset in an exhaustive manner.
This means that DataSite's scheduling algorithm requires fine tuning; pure brute force, as somewhat provocatively stated earlier in this paper, is not a universal solution.
Our current implementation can scale up to tens of thousands of entries in the dataset, which is comparable to many existing visualization tools~\cite{Wongsuphasawat2017, Yalcin2017}.
In particular, our evaluations involve datasets with 10,000 (bird strikes) and 3,000 (movies) items.
Still, while there is potential for a more scalable system, it is beyond the scope of this paper.

\subsection{Top-Down vs.\ Bottom-Up Data Analysis}

One of the strengths of visualization is its data-driven, bottom-up, and self-informing nature: as Tukey notes~\cite{tukey1977exploratory}, the type of \textit{exploratory data analysis} so powerfully supported by visualization allows for deriving hypotheses and insights from datasets that are not previously known or well-explored.
This same focus on hypothesis generation permeates much research on visual exploration, including in particular Keim's seminal work~\cite{Keim2001}, which quotes visualization as ``especially useful when little is known about the data and the exploration goals are vague.''
Put simply, visualization allows you to ask (and often answer) questions you didn't know you had.
This is also the strength of a visualization recommendation engine such as Voyager 2, where the philosophy can be expressed as generating as many pertinent charts as possible in the hope of informing the user.

It is also diametrically opposite to the top-down approach afforded by the server-side computation engine used in DataSite, where a suite of pre-defined computational modules are used to extract potentially significant features from a dataset and bring them to the user's attention.
The significant difference between this and traditional confirmatory data analysis methods, including statistical packages such as R, SAS, and JMP, is that DataSite eliminates the need for both (a) forming hypotheses, and (b) testing them using the correct methods.
It does this in the most possible straightforward way: by relying on sheer brute force to test \textbf{all} the hypotheses through state-of-the-art computational modules designed by the DataSite developers.
However, by definition, such a suite of modules is limited by the actual modules provided, which makes this approach less flexible to unknown datasets.
A bottom-up visualization-centric approach, on the other hand, will rely on the human user to detect incidental features in the dataset.
This means that DataSite trades some of the flexibility of more open-ended visual exploration tools for the benefit of reducing the knowledge and hypothesis generation barriers of such tools.

It is important to keep this trade-off in mind when contrasting top-down vs.\ bottom-up data analysis tools, such as those compared in this paper.
The ultimate purpose of this paper is to explore this trade-off in more detail, not to attempt to demonstrate the superiority of one approach or the other.
No approach is likely to be superior, and, in fact, their combination will likely reap the most rewards.
For example, our DataSite implementation also includes manual view specification to enable the user to independently visualize the data, and also uses charts even when reporting on features found.
This is done in an effort to stimulate the type of serendipitous, bottom-up sensemaking that visualization scaffolds.
It is clear that such a combined effort is the best way to proceed in this domain even in the future.

\section{Conclusion and Future Work}
\label{sec:conclusion}

We have presented DataSite, a visual analytics system that integrates automatic computation with manual visualization exploration.
DataSite introduces the feed, a list of dynamically updated notifications arising from a server-side computation engine that continually runs suitable analyses on the dataset. 
The feed stimulates the analyst's sensemaking through brief descriptions of computational modules along with corresponding charts.
Filters and text search bar enable quick scan and fast data exploration.
Two controlled user studies evaluate the approach compared to PoleStar and Voyager 2, respectively, and show that significant performance improvements over the manual view specification tool (PoleStar) in both breadth and depth for data coverage, as well as useful guidance in exploration.
It also provides more meaningful charts and features to analysts over Voyager 2, while maintaining similar ease of usage.
The results are promising and indicate that the system promotes data analysis in all stages of exploration. 

DataSite can be seen as a canonical visual analytics system in that it blends automatic computations with manual visual exploration, thus establishing a true partnership between the analyst and the computer.
We regard it as the first step towards a fully proactive visualization system involving a human in the loop. 
Of course, many improvements can be made towards a more efficient system; after all, while CPU resources are cheap, they are not free.
One potential future research topic is guiding recommendations based on the analyst's interest, past interactions, and even their personality. 
For example, consider a DataSite-like system that would respond to an analyst drilling deep into a part of a sales dataset over time to proactively compute future sales projections for that part of the data in an effort to anticipate future questions the analyst may have.
Other ideas may include mining the analyst's click stream, browsing and analysis history, and even social media profiles to determine how to best guide the proactive computation.
Finally, we could also use interaction to dynamically update the ranking of features in the feed, e.g., prioritize features for data fields selected by the user.

\section*{Acknowledgments}

This work was partially supported by the U.S.\ National Science Foundation award IIS-1539534.
Any opinions, findings, and conclusions expressed in this material are those of the authors and do not necessarily reflect the views of the funding agency.

\bibliographystyle{SageH}
\bibliography{datasite}

\end{document}